\documentclass[10pt]{article} 
\usepackage[preprint]{tmlr}

\usepackage{hyperref}
\usepackage{url}
\usepackage{microtype}
\usepackage{mathtools}
\usepackage{amssymb}
\usepackage{booktabs}
\usepackage{tikz}
\usepackage{amsthm}
\usepackage{graphicx}
\usepackage{amsmath}
\usepackage{natbib}

\usepackage{amsmath,amsfonts,bm}
\usepackage[T1]{fontenc}
\usepackage{algorithm,algpseudocode}
\usepackage{graphicx}
\usepackage{enumitem}
\usepackage{wrapfig}
\usepackage{xfrac}
\usepackage{subcaption}
\usepackage{longtable}
\usepackage{array}
\usepackage{xcolor}

\DeclareMathOperator*{\argmin}{arg\,min}

\usepackage{amsthm}

\theoremstyle{definition}

\theoremstyle{plain}

\title{Completion of partial structures using Patterson maps with the CrysFormer machine learning model}

\author{
   \name Tom Pan \email qp3@rice.edu \\
   \addr Department of Computer Science, Rice University, Houston, TX 77005, USA 
   \AND
   \name Evan Dramko \email ed55@rice.edu \\
   \addr Department of Computer Science, Rice University, Houston, TX 77005, USA 
   \AND
   \name Mitchell D. Miller \email mitchm@rice.edu \\
   \addr Computer Science Dept., Rice University, Houston, TX, USA
   \AND
   \name Anastasios Kyrillidis \email anastasios@rice.edu \\
   \addr Department of Computer Science, Rice University, Houston, TX 77005, USA \\
          Ken Kennedy Institute, Rice University, Houston, TX 77005, USA
   \AND
   \name George N. Phillips, Jr. \email georgep@rice.edu \\
   \addr Department of BioSciences and Department of Chemistry, Rice University, Houston, TX 77005, USA
}

\begin{document}
\maketitle
\begin{abstract}
Protein structure determination has long been one of the primary challenges of structural biology, to which deep machine learning (ML)-based approaches have increasingly been applied.
However, these ML models generally do not incorporate the experimental measurements directly, such as X-ray crystallographic diffraction data. 
To this end, we explore an approach that more tightly couples these traditional crystallographic and recent ML-based methods, by training a hybrid 3-d vision transformer and convolutional network on inputs from both domains.
We make use of two distinct input constructs / Patterson maps, which are directly obtainable from crystallographic data, and ``partial structure'' template maps derived from predicted structures deposited in the AlphaFold Protein Structure Database with subsequently omitted residues. 
With these, we predict electron density maps that are then post-processed into atomic models through standard crystallographic refinement processes.
Introducing an initial dataset of small protein fragments taken from Protein Data Bank entries and placing them in hypothetical crystal settings, we demonstrate that our method is effective at both improving the phases of the crystallographic structure factors and completing the regions missing from partial structure templates, as well as improving the agreement of the electron density maps with the ground truth atomic structures.
\end{abstract}

\textit{This work has been accepted in Acta Crystallographic section D.}

\section{Introduction}
\label{sec:introduction}

Proteins are essential components of nearly all biochemical mechanisms performed in living cells \citep{tanford2004}. 
They are composed of small organic molecules called amino acids (of which there are 20 typical proteinogenic ones) linked by peptide bonds; a single amino acid is often referred to as a residue.
Proteins’ functions are largely facilitated by their ability to bind only to specific molecules at specific sites on the protein, such that its 3-D shape significantly informs its cellular activity. 
Thus, determining the characteristic complex 3D structure of a protein (which had folded from a polymer of amino acid residues) is a long-standing problem of structural biology, having first been achieved by X-ray crystallography, and later by Nuclear Magnetic Resonance (NMR) and cryoelectronmicroscopy (cryo-EM). 
All of these approaches face the problem of reconstructing an atomic structure given incomplete or imperfect experimental data \citep{drenth2007principles}. 
In recent years, due to the wealth of high-quality information that has been accumulated in the Protein Data Bank (PDB) and vast sequence databases, machine learning has become another widespread approach for predicting protein structure, often with model architectures based on the transformer self-attention mechanism.
Initiatives such as AlphaFold2 \citep{jumper2021highly} and AlphaFold3 \citep{AF3}, which use sequence data in conjunction with co-evolutionary information in the form of multiple sequence alignments (MSAs), have demonstrated the ability of deep learning models to produce variously precise atomic-level predictions.
Other established ML-based approaches include RoseTTAFold \citep{rosettafold}, Boltz \citep{Boltz-1} and ESMFold \citep{esmfold}, which does not depend on the creation of MSAs.
However, certain issues remain (Terwilliger et al., 2023b), and X-ray crystallography is still frequently employed despite its well-known associated difficulties (i.e. the crystallographic phase problem).  

Thus, several projects have been developed with the purpose of bridging the gap between experimental crystallographic methods and ML techniques, see \citep{2024_Matinyan}.
In this work, we build upon our previous work \citep{pan2023deep,pan2025reccrysformer} in this area by developing an ML-based approach for improving predictions provided by other ML models, i.e. those in the AlphaFold Protein Structure Database (AFDB) \citep{afdb2024}, given Patterson maps, which can be directly calculated from X-ray crystallography diffraction patterns without the need for phase information.
We develop a novel synthetic dataset of crystals of protein segments taken from PDB structures and corresponding AFDB entries, with a subset of residues subsequently omitted from the AFDB prediction templates, and show that a hybrid 3D vision transformer and convolutional neural network (CNN) can be trained to complete and improve the templates extracted the AlphaFold predictions.

\section{Problem Setup and Related Work}
\label{sec:definition}

\subsection{X-ray crystallography.} 
X-ray crystallography is one of the most commonly used experimental methods for determining the atomic-level structures of proteins and other large macromolecules\citep{lattman2008}. 
In this technique, molecular crystals are exposed to X-ray beams, which diffract in specific directions according to the regular internal structure of the crystal to produce a pattern of spots (known as reflections).

Each reflection with a crystallographic diffraction pattern is associated with a set of three Miller indices $h, k, l$ that indicate the orientation of sets of parallel planes within the crystal’s unit cell that contribute to producing the reflection \citep{ashcroft2022solid}, and can be shown to have an underlying representation known as a structure factor.
Formally, a structure factor is the Fourier transformation of the electron density within the unit cell (the smallest repeating unit within a crystal).
However, it can be well-approximated as a discrete Fourier transform dependent on the atoms present in the crystal's unit cell: \vspace{-0.2cm}
\begin{align}
F(h,k,l) = \sum_{j=1}^{n} f_j \cdot e^{2\pi i(hx_j + ky_j + lz_j)} \cdot e^{-\tfrac{B_j}{4} (\tfrac{1}{d_{hkl}})^2}, \\[-20pt] \nonumber
\vspace{-0.8cm}
\end{align}
where $f_j$ refers to the scattering factor property, $B_j$ refers to the crystallographic B-factor, $(x_j, y_j, z_j)$ refers to the fractional coordinates of the $j$-th atom within the cell (and all occupancies are assumed to be $1.0$).
Each of these structure factors is known to be a complex number, with both an amplitude and a phase (denoted by $\phi$) component.
An inverse Fourier transform can be taken over all such reflection structure factors to obtain an initial estimate of the electron density $\rho$ at all points $(x,y,z)$ within the unit cell, as: \vspace{-0.2cm}
\begin{align}
\rho(x,y,z) = \tfrac{1}{V} \cdot \sum_{h,k,l}^{} |F(h,k,l)| \cdot e^{-2\pi i(hx + ky + lz - \phi(h,k,l))}, \\[-20pt] \nonumber
\vspace{-0.2cm}
\end{align}
where $V$ is the volume of the unit cell. 
Once such a map of the electron density within the unit cell is estimated, it is used to produce an initial input into iterative refinement programs, which perform repeated comparisons of the expected diffraction pattern given the current estimated model with experimental measurements, and eventually output a final atomic model.
However, while the amplitude $|F(h,k,l)|$ of any reflection's underlying structure factor is simply proportional to the square root of its measured intensity, corresponding phase information cannot be immediately calculated from experimental crystallographic data. This is known as the crystallographic phase problem \citep{lattman2008}.

\subsection{Previous work for solving the crystallographic phase problem.} 
Traditionally, three of the most widely used methods for addressing the crystallographic phase problem have been isomorphous replacement (IR), anomalous dispersion (AD), and molecular replacement (MR) \citep{lattman2008, jin2020molecular}. 
IR and AD almost always require multiple experimental settings, often with the production of molecular crystals with heavy atom substitutions.
On the other hand, MR requires the availability of homologous structures known to be similar to the current desired one, to be used as an initial template or phase estimate after a rotational and then translational search.
Predictions made by the AlphaFold2 machine learning model \citep{jumper2021highly} have been effectively used as initial models for the MR technique \citep{McCoy:qg5003}, especially as part of an iterative process akin to traditional crystallographic refinement involving multiple rounds of MR and model building \citep{Terwilliger:nz5011}.
Furthermore, in our previous work \citep{pan_sdy}, we obtained essentially \textit{de novo} structural predictions of short protein fragments from corresponding Patterson maps using the \texttt{CrysFormer} model architecture.
Our current work now aims to show the viability of incorporating an additional machine learning step, that makes use of our established \texttt{CrysFormer} model, to improve existing AlphaFold2 predictions given crystallographic data, representing a further integration of experimental and ML-based protein structure determination methods.

\subsection{The Patterson function.}
The \textit{Patterson function} \citep{Patterson} is often used as an intermediary during the aforementioned methods for solving the crystallographic phase problem. 
It is a variation of the Fourier transform from structure factors to electron density where the amplitude components are squared and phases are ignored, resulting in what is called a Patterson map: \vspace{-0.2cm}
\begin{align}
p(u,v,w) = \tfrac{1}{V} \cdot \sum_{h,k,l}^{} |F(h,k,l)|^{2} \cdot e^{-2\pi i(hu + kv + lw)}.\\[-18pt] \nonumber
\end{align}
where $(u, v, w)$ refers to locations within the Patterson map’s unit cell, which has the same dimensions as that of the original crystal.
Since phase information is not needed, Patterson maps can be immediately computed from raw crystallographic experimental data.
And if an underlying (real-valued, real-space) electron density map is denoted as $\mathbf{e} \in \mathbb{R}^{N_1 \times N_2 \times N_3}$, then the corresponding Patterson map $\mathbf{p}$ can alternatively be formulated as follows:
\begin{align}
    \mathbf{p} = \Re\left(\mathcal{F}^{-1} \left( \mathcal{F}(\mathbf{e}) \odot \mathcal{F}(\widehat{\mathbf{e}})\right) \right) = \Re\left(\mathcal{F}^{-1} \left( |\mathcal{F}(\mathbf{e})|^2\right) \right),
\end{align}
where $\odot$ refers to element-wise multiplication, $\mathcal{F}$ refers to the Fourier transform, and $\Re$ emphasizes that the result is a real number. 
$\widehat{\mathbf{e}}$ refers to an inverse-shifted version of $\mathbf{e}$, where each entry is defined as $\widehat{e}_{i, j, k} = e_{N_1 - i, N_2 - j, N_3 - k}$.

From the construction of the Patterson function, a further derivation indicates that a Patterson map does not directly reveal the atomic structure within a unit cell.
Instead, (disregarding thermal motion effects and assuming infinite resolution) each peak in a Patterson map essentially corresponds to an interatomic vector between atoms within the crystal's unit cell, and so Patterson maps of large macromolecules such as proteins are extremely dense with peaks (the amount of which scales quadratically with the number of atoms in the original cell), that may often blur together.
Also, the height of these peaks is proportional to the product of atomic numbers in the corresponding pair (or the sum of all such pairs that have identical interatomic vectors), allowing the contributions of heavier atoms to dominate the resulting map.
This is actually desired if they were explicitly incorporated or substituted into the molecular structure as in IR or AD.
These issues prevent the straightforward interpretation of crystallographic Patterson maps, and so they have not been used to directly estimate the corresponding electron densities.

\section{Model Completion with Partial Structure Inputs}
\label{sec:method}

\subsection{Using deep learning.}
Patterson maps can be used as inputs into machine learning models, as constructs directly obtainable from raw crystallographic data without additional experiments or outside information.
Thus our goal is to train a model $g$ with parameters $\boldsymbol{\theta}$ to estimate electron density maps given corresponding Patterson maps as input (see \ref{Model architecture}).
Formally, given a dataset of $n$ examples of the form $(\mathbf{p}_i, \mathbf{e}_i)_{i = 1}^n $, where $\mathbf{p}_i \in \mathbb{R}^{N_1 \times N_2 \times N_3}$ is the Patterson map that corresponds to a ground truth electron density map, $\mathbf{e}_{i} \in \mathbb{R}^{N_1 \times N_2 \times N_3}$, we aim to obtain optimal model parameters $\boldsymbol{\theta}^\star$ such that our model predictions are as close as possible to the ground truth maps given a loss function $\mathcal{L}(\boldsymbol{\theta})$:


\begin{align}
\boldsymbol{\theta}^\star &= \underset{\boldsymbol{\theta} }{\argmin} ~\left\{ \mathcal{L}(\boldsymbol{\theta}) := \tfrac{1}{n}\sum_{i = 1}^n \ell(g(\boldsymbol{\theta}, \mathbf{p}_i), \mathbf{e}_i) \right\} \\[-10pt] \nonumber
\end{align}

We use mean squared error (MSE), well-established for regression tasks, as our primary internal loss function. 
However, we also take the negative Pearson correlation between ground truth and predicted maps as an additional loss function term.
This comparison between two constructs of the same shape is used across a wide range of application domains, including crystallography. 
Denoting a model prediction as $\mathbf{e'}$, and defining the average value over a ground truth map and predicted map as $\bar{\mathbf{e}}=\tfrac{1}{N_1N_2N_3}\sum_{i,j,k}\mathbf{e}_{i,j,k}$ and $\bar{\mathbf{e}}'=\tfrac{1}{N_1N_2N_3}\sum_{i,j,k}\mathbf{e}'_{i,j,k}$, respectively, the Pearson correlation coefficient is defined as:
\begin{small}
\begin{align}
\texttt{PC}(\mathbf{e},\mathbf{e}') = 
\frac{\sum\limits_{i,j,k = 1}^{N_1,N_2,N_3} (\mathbf{e}'_{i,j,k} - \bar{\mathbf{e}}') (\mathbf{e}_{i,j,k} - \bar{\mathbf{e}})}{\rule{0pt}{1.6em} \sqrt{\sum\limits_{i,j,k = 1}^{N_1,N_2,N_3} (\mathbf{e}'_{i,j,k} - \bar{\mathbf{e}}')^2} \cdot \: \sqrt{\sum\limits_{i,j,k = 1}^{N_1,N_2,N_3} (\mathbf{e}_{i,j,k} - \bar{\mathbf{e}})^2}},
\end{align}
\end{small}

As larger Pearson correlations indicate greater agreement between maps, we negate the calculated values to incorporate them into our overall loss function.

\subsection{Using existing predictions as partial structure templates.} 

In our previous work \citep{pan_sdy}, the only additional input information provided to the model beyond Patterson maps was in the form of electron density maps corresponding to single amino acid residues in their most common conformations, which were referred to as "partial structures".
But for the current problem, we instead make use of existing predictions obtained from the AFDB as our partial structures. 
We omit a subset of residues from these existing predictions to simulate realistic conditions, where often portions of a protein structure prediction from an ML model would have regions of low confidence and accuracy (which we aim to fill in using experimental crystallographic data).
We train our model to complete and improve these initial templates provided by the incomplete AlphaFold predictions, and thus no longer directly determine protein structures solely from crystallographic data, but instead incorporate both existing experimental and machine learning approaches for structural prediction into a unified framework.
We now aim to optimize:

\begin{align}
\boldsymbol{\theta}^\star &= \underset{\boldsymbol{\theta} }{\argmin} ~\left\{ \mathcal{L}(\boldsymbol{\theta}) := \tfrac{1}{n*J}\sum_{i = 1}^n \sum_{j = 1}^J \ell(\boldsymbol{\theta};~g, (\mathbf{p}_i, \mathbf{e}_i, \mathbf{u}^{j}_{i})) \right\}. \\[-20pt] \nonumber
\end{align}

where each original Patterson map and ground truth pair $(\mathbf{p}_i, \mathbf{e}_i)$ is associated with multiple different corresponding incomplete template "partial structures" $\mathbf{u}^{j}_i$, with each of these (in practice up to) $J$ partial structures having a different subset of removed residues.
The full dataset size is denoted as $n*J$ in the equation for simplicity, but in practice is slightly smaller than this.

\subsection{Model architecture} \label{Model architecture}

\begin{wrapfigure}{r}{0.35\textwidth}
    \vspace{-.5cm}
    \begin{minipage}{0.35\textwidth}
        \includegraphics[width=1\linewidth]{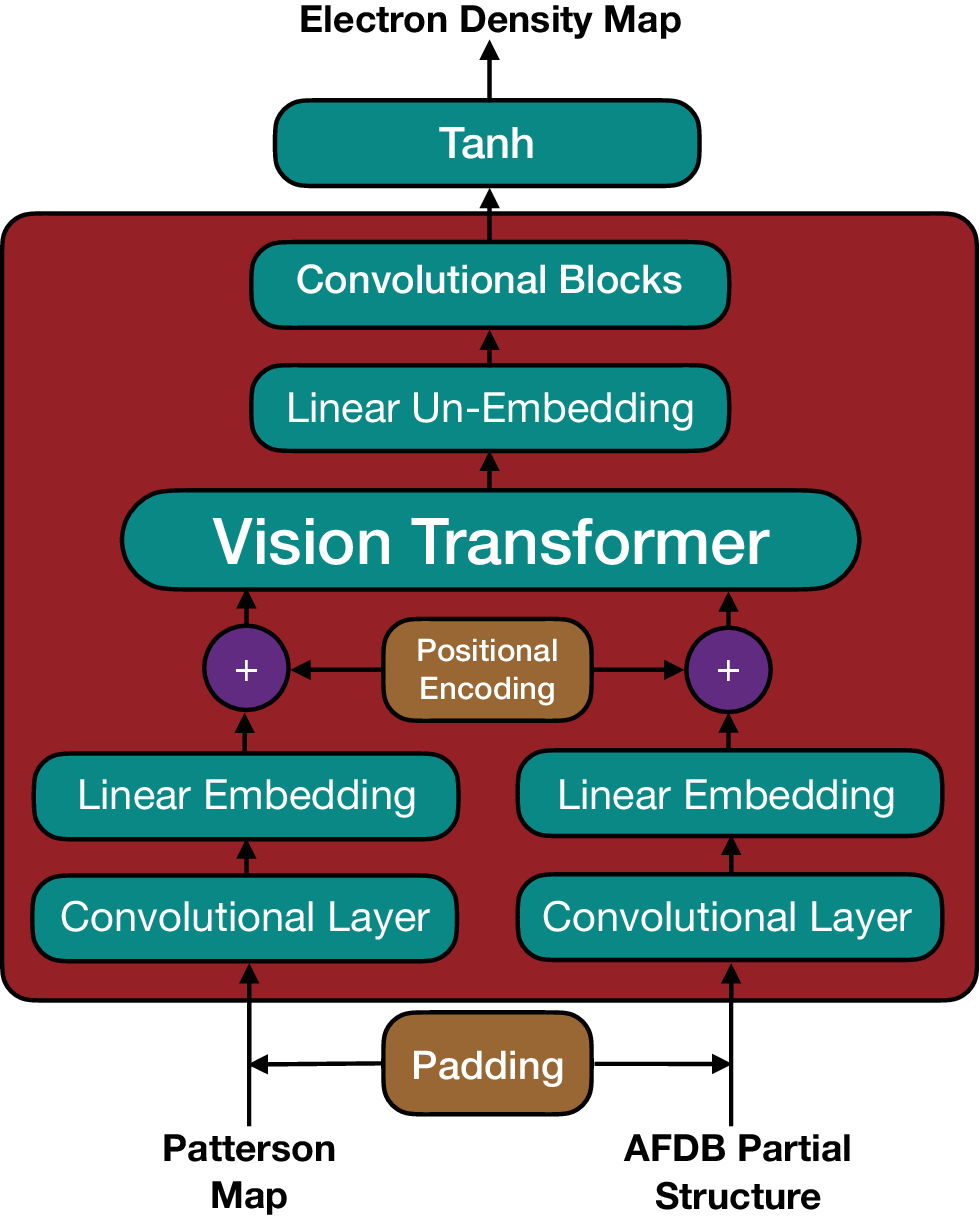} 
    \end{minipage}
    \caption{Overview of \texttt{CrysFormer} model architecture }
    \label{fig:rec} \vspace{-0.95cm}
\end{wrapfigure}

As stated, we continue to use the \texttt{CrysFormer} \citep{pan_sdy} model introduced in our previous work for the model completion task.
This model is a hybrid of a 3d vision transformer and CNN, with Nyström approximate attention \citep{Xiong_Zeng_Chakraborty_Tan_Fung_Li_Singh_2021} in the self-attention layers of the transformer.
For this work, we begin to use scale-equivariant 3d convolution and batch normalization layers introduced by \citep{wimmer2023scaleequivariantdeeplearning3d} in all such layers before the transformer.
Also, we no longer provide several partial structure templates for each dataset example, each of a smaller size than the corresponding Patterson map input, but instead provide one single partial structure of the exact same size as the Patterson and desired ground truth maps.
Furthermore, every Patterson and ground truth pair now corresponds to up to $J$ distinct dataset examples (Fig. \ref{fig:rec}).

\section{Dataset Generation}
\label{sec:dataset}

\begin{wrapfigure}{l}{0.43\textwidth}
    \vspace{-0.15cm}
    \begin{minipage}{0.43\textwidth}
        \includegraphics[width=1\linewidth]{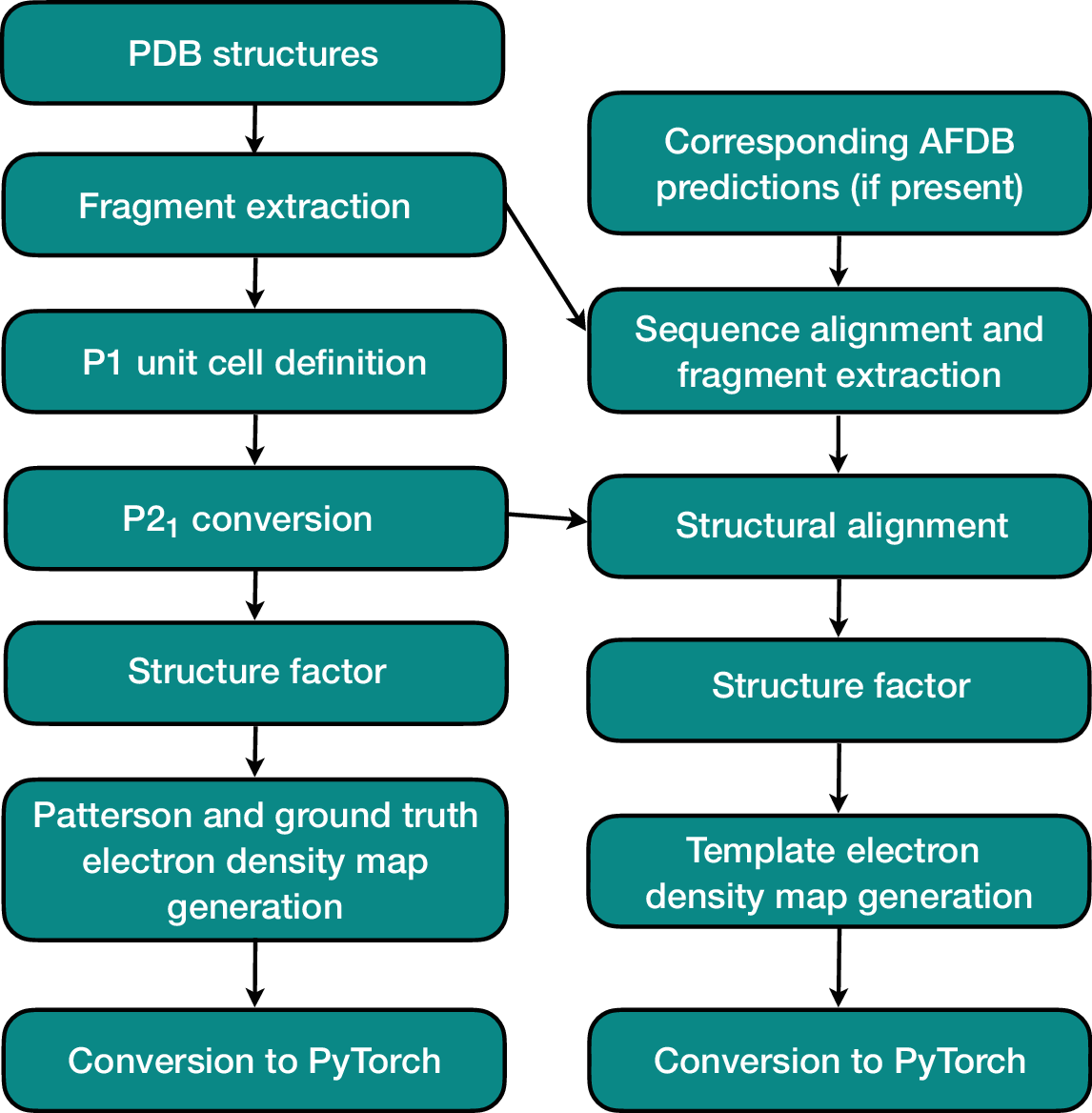} 
    \end{minipage}
    \caption{High-level steps of our dataset generation process}
    \label{fig:rec2} \vspace{-0.75cm}
\end{wrapfigure}

We followed the same overall data generation process as described in our previous work\citep{pan_sdy}, but with several modifications to better fit the new task of model completion on a single input partial structure (Fig. \ref{fig:rec2}).
We started with an expanded initial basis of nearly $38,000$ PDB protein structures, curated according to the following criteria kept as before:
Solved by X-ray crystallography between the years 1995-2023, with sequence length $\geq$ 40, refinement resolution $\leq$ 2.75 {\AA}ngstr\"{o}ms ({\AA}), R-Free $\leq$ 0.28, and available in legacy PDB format. 
As we desired a much larger dataset for the current problem than that of our previous reported work, we increased the clustering sequence identity criterion to $70\%$ from $30\%$ to increase the size of the starting basis, and extracted all possible 15-residue fragments without randomly removing any obtained ones.
Further, we allowed overlaps of up to 5 out of 15 residues between consecutive extracted fragments.
Thus we continued to place all examples derived from the same initial protein structure together in either the training or test set, preventing our model from potentially simply memorizing regions of protein segments that are present in the training set when evaluating unseen examples.

\begin{figure}[!t] 
    \centering
    \includegraphics[width=0.23\textwidth]{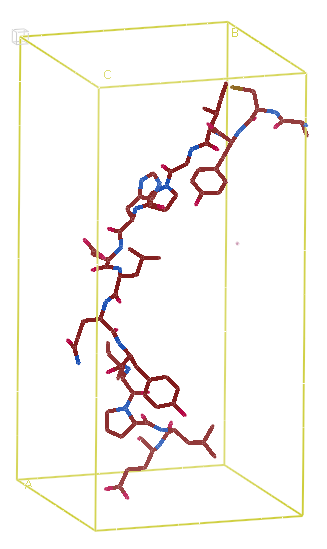} \hspace{0.3cm}
    \includegraphics[width=0.4\textwidth]{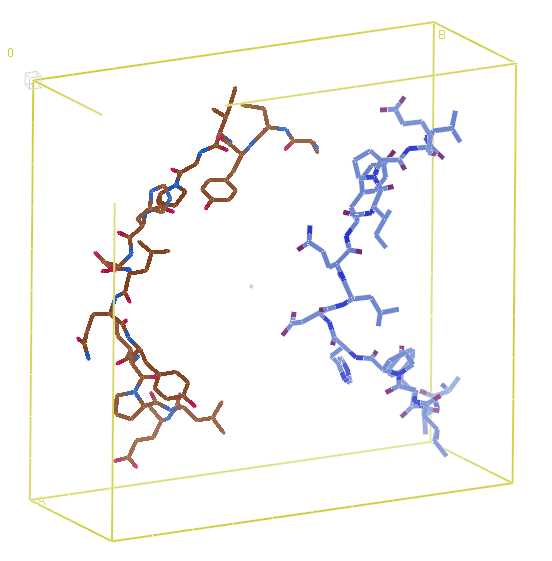} \hfill
\caption{A 15-residue protein fragment extracted from a PDB structure, in a reoriented P1 unit cell (left) and after conversion to P2$_1$ (right).}
\label{fig:p21}
\end{figure}

We applied most of our previous standardized modifications to these protein fragments, such as removing examples containing nonstandard or missing residues or missing atoms using the \texttt{pdbfixer} \citep{eastman2017openmm} Python library, removing all Hydrogen atoms, and converting Selenomethionine residues to Methionine. 
As an additional form of variability, we did not reset all atomic temperature factors to a constant value, but instead kept all such values from the original PDB structure.
We determined the original unit cell extents for our fragments starting from the raw $\max-\min$ ranges of Cartesian coordinates along each of the three axes, iteratively increasing the current dimensions until the minimum intermolecular atomic contact was at least $3.5$ {\AA}.
We converted our examples to the P2$_1$ space group, which is one of the most common space groups (such unit cells contain two molecules related by a screw axis).
First, we reoriented all initial unit cells so that the first axis is the longest and the second axis (along which the P2$_1$ screw axis is located by convention) is the shortest.
Then when converting to P2$_1$, we added an additional {\AA}ngstr\"{o}m to each original dimension and further expanded the length of the second axis by a multiplier randomly selected from the range $1.7-1.95$.
Another key consideration for this dataset was obtaining a much more realistic solvent content (and thus the amount of empty space) within the unit cell compared to our previous dataset, so we did not check for and remove examples that no longer satisfied the minimum $3.5$ {\AA} atomic contact requirement (Fig. \ref{fig:p21}).
However, for each example, we still centered atomic coordinates such that the center of mass was at the unit cell's exact center to avoid ambiguities associated with the translation invariance of Patterson maps; this is theoretically justifiable as unit cell boundaries relative to the contents thereof are essentially arbitrary.

We again generated structure factors for each example in its final P2$_1$ unit cell with the \texttt{gemmi sfcalc} program \citep{wojdyr2022gemmi}, without any bulk solvent scaling or correction, and then created both Patterson and ground truth electron density maps in the .ccp4 format from these structure factors with the \texttt{fft} program of the \texttt{CCP4} program suite \citep{CCP4:2023}.
We divided the dataset as evenly as possible into 20 bins, each assigned to a different resolution limit in the range $1.75$ {\AA} to $2.3$ {\AA}, and restricted the structure factors used to generate the maps to the corresponding specified resolution limit.
We also associated each of the resolution limit bins with a different grid sampling factor in the range of $2.29$ to $2.7$.
We divided this value by the corresponding selected resolution limit, and used the result as a multiplier on the P2$_1$ unit cell dimensions when specifying the map dimensions.
As before, the Patterson and electron density maps for each example had the exact same dimensions and resolution limits.
The Patterson and ground truth electron density maps were then converted into PyTorch tensor format, maintaining the map axis dimensions, and all values in each tensor were normalized according to the maximum and minimum element values for the corresponding map type over the entire dataset to be in the range $[-1, ~1]$.
The maximum and minimum values were found separately for the set of Patterson and electron density maps, as Patterson maps are far more dense than electron densities.

To create our partial structure templates, we first queried the PDBe SIFTS database \citep{pdbe, siftsDB} to associate UniProt IDs with each of our original protein structures, given their PDB and entity IDs. 
For all structures with an associated UniProt ID, we obtained the associated AlphaFold ID if present in the accession\_ids.csv file provided in the full AlphaFold Database. 
If such an AlphaFold ID was found, we then downloaded the v4 version of the corresponding AlphaFold prediction from the AFDB \citep{afdb2024}.  Structures without an AFDB match were excluded from the dataset to avoid the computational overhead of running AlphaFold to generate suitable predicted models.
We used the Needleman-Wunsch sequence alignment method \citep{Needleman-Wunsch}, implemented as a Java program \citep{nwalign}, to align each of our protein fragment examples to the AlphaFold prediction corresponding to the original structure it was extracted from. 
Using these alignments, we extracted all segments corresponding to our examples from the corresponding AlphaFold prediction coordinate files.
After obtaining AlphaFold fragments for each remaining example, we applied the same set of standardized modifications as we had to the original fragments, although we reset all temperature factors for the atoms in the AlphaFold structures to a constant $20.0$.
We then performed a structural alignment of each AlphaFold segment to its corresponding ground truth segment. For computational expediency in generating the dataset, we used the align command in PyMOL \citep{PyMOL} (matching the backbone and C$_\beta$ atoms) as a proxy for fragment placement with an MR program (like phaser \citep{phaser}). We saved the aligned AlphaFold segment in a unit cell that matched the corresponding ground truth coordinate file.

\begin{figure}[!t] 
    \centering
    \includegraphics[width=0.4\textwidth]{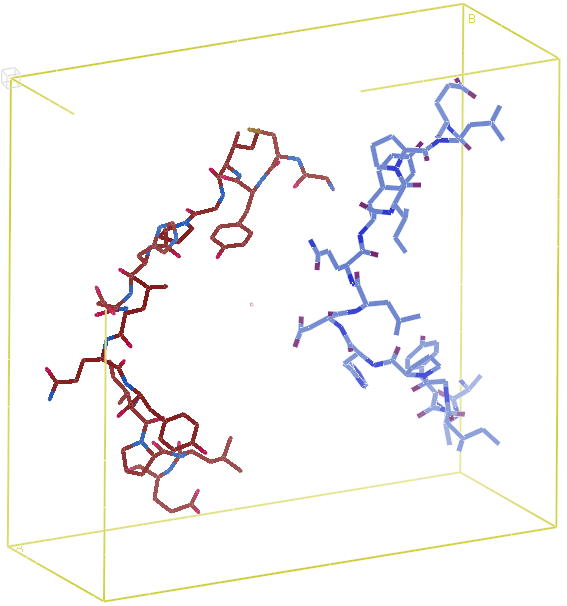} \hspace{0.3cm}
    \includegraphics[width=0.4\textwidth]{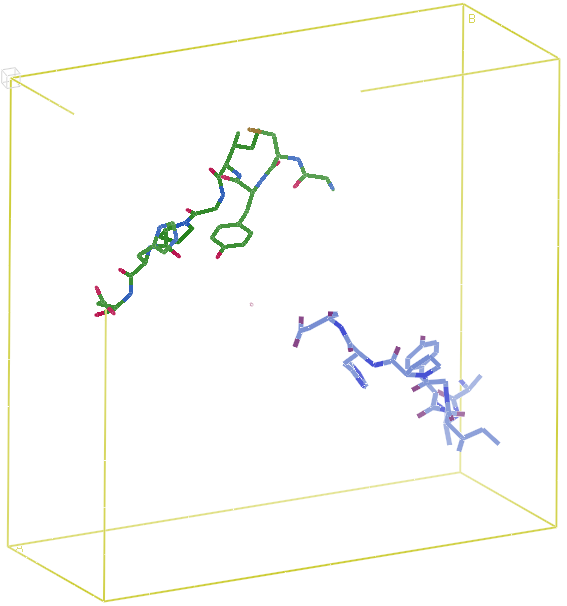} \hfill
\caption{A 15-residue protein fragment extracted from a predicted structure in AFDB, placed in a P2$_1$ unit cell (left) and a partial structure with a random amount of residues omitted from a random end of the extracted fragment (right).}
\label{fig:ps}
\end{figure}

We then removed a subset of the residues from the AlphaFold fragments via a sequence of 2-3 random selections (Fig. \ref{fig:ps}).
For each fragment, we generated up to three such partial structures with omitted residues.
When removing residues, we first randomly selected whether to begin from the start of the coordinate file, the end of the file, or both ends.
Then, we randomly selected the number of residues to omit from $3-7$ out of $15$ total.
If only one end to omit from was selected, we removed the selected number of residues contiguously.
If both ends were selected, we omitted half of the total selected number from one end and the rest from the other end (if an odd number of residues was to be omitted, we simply performed another random selection to determine which end to remove more residues from).
It is possible that the exact same end and number of omitted residues was selected as that of a previously created partial structure for a particular fragment.
If this occurred, we did not try to create another partial structure, but simply allowed for fewer than $J$ partial structures to be associated with that fragment (thus the total dataset size was slightly smaller than $n*J$).
Afterward, we applied the rest of our dataset generation process to the partial structures, but only needed to generate electron density maps and not Patterson maps as well.
For each partial structure, we specified the same map dimensions and resolution limit as the corresponding ground truth fragment. 
And when performing max-min normalization on the partial structure tensors, we used the exact same maximum and minimum as we had when normalizing the ground truth electron density tensors.
Finally, to ensure uniform shape across the examples in each of our training batches as required by PyTorch, examples that belonged to tensor-size bins smaller than our minimum batch size of 6 were again excluded from the training set.
This was far less likely than before as every ground truth map in the training set is now associated with up to three distinct partial structures.

\section{Experiments}
\label{sec:results}

\subsection{Training Details}
We performed a single training run of our model on a training set of $589,546$ initial Patterson Map-ground truth electron density pairs, where each original example was associated with up to three distinct "partial structures" with $3-7$ subsequently omitted residues out of $15$ as described above ($J$ = 3), for a total size of $1,634,839$ examples.
These were split into batches of minimum size $6$, average size $10$, and maximum size $11$. 
For training, we used a "Schedule-Free" variant of the AdamW optimizer \citep{defazio2024road}, although we still enforced an overall OneCycle learning rate schedule \citep{Smith19}. 
The model was trained for 71 epochs in a data-parallel fashion using PyTorch's DDP module \citep{ddp2020} on a pair of RTX 6000 Ada GPUs with 48 GiB memory each, with \texttt{torch.set\_float32\_matmul\_precision} set to \texttt{'high'} and gradient accumulation performed every 2 batches.
We report the hyperparameters of our model architecture used for this training run in Table \ref{hyper}. 
Furthermore, we downsample our structures by the patch size of $4$ in every axis after the 2nd and 4th transformer layers with a 3d convolution, and upsample by the same amount after the 8th and 10th transformer layers with a 3d transposed convolution. 

\begin{table*}[!htp]
\caption{Hyperparameter settings used in training run.}
\label{hyper}
\centering
\begin{tabular}{lc}
\toprule
    Hyperparameter  &  Value \\ 
    \midrule
    Convolution output channels & 10 \\
    Patch size & 4x4x4 \\
    Embedding dimension & 512 \\
    Head dimension & 64 \\
    Number of heads & 12 \\
    MLP dimension & 2048 \\
    Transformer layers & 12 \\
    AdamW weight decay & $3e-2$ \\
    inital lr & $4.5e-4$ \\
    maximum lr & $2.85e-3$ \\
    final lr & $8.57e-4$ \\
    \bottomrule
\end{tabular}
\end{table*}

\subsection{Metrics}
As a baseline comparison for our model predictions after training, we use the \texttt{sigmaa} program of the \texttt{CCP4} program suite \citep{CCP4:2023} with the "PARTIAL" option specified to improve the maps derived from partial structures with removed residues. 
Resolution ranges were specified to be the same as those used to generate .ccp4 maps from structure factors during our dataset generation process.
For evaluation, we use the \texttt{get\_cc\_mtz\_pdb} program of the \texttt{Phenix} program suite \citep{Liebschner2019:di5033} to calculate the Pearson correlation coefficient, as defined previously, passing either post-\texttt{sigmaa} structure factors or model prediction-derived structure factors without \texttt{sigmaa} weighting, and the corresponding ground truth atomic coordinates. 
We report the Pearson correlations calculated only over the region of the unit cell where the atomic model is located, ignoring empty regions.
Additionally, we perform phase error analysis using the \texttt{cphasematch} program \citep{cphasematch}, again of the \texttt{CCP4} program suite.  
We report both unweighted and FOM-weighted average phase errors in degrees, where a smaller phase error is desirable, in Table \ref{phase_errors}.

\begin{table*}[!htp]
\caption{ Comparison of structure factors derived from model predictions versus structure factors of partial structure templates after applying \texttt{sigmaa}. Metrics averaged over test set examples or a subset thereof. Standard deviations reported in parentheses after mean values.} 
\label{phase_errors}
\centering
\begin{tabular}{lcccc}
\toprule
               & Predictions  & Sigmaa  & Predictions  & Sigmaa  \\ 
    Metric    & (Full) & (Full) & (Subset) & (Subset) \\  \midrule
    Phase error (unweighted; \textdegree) & \textbf{39.4} (10.3) & 48.7 (10.5) & \textbf{52.7} (14.4) & 62.6 (13.3)  \\
    Phase error (unweighted; cosine) & \textbf{0.762} (0.130) & 0.649 (0.146) & \textbf{0.589} (0.209) & 0.448 (0.205)  \\
    Phase error (weighted; \textdegree) & \textbf{30.2} (8.4) & 38.5 (10.0) & \textbf{41.2} (13.4) & 52.2 (14.7)  \\
    Phase error (weighted; cosine) & \textbf{0.856} (0.093) & 0.772 (0.127) & \textbf{0.734} (0.177) & 0.594 (0.212)  \\
    Pearson correlation coefficient & \textbf{0.867} (0.074) & 0.816 (0.100) & \textbf{0.783} (0.136) & 0.692 (0.171)  \\
    \bottomrule
\end{tabular}
\end{table*}

\subsection{Results and comparison with baseline}
We provide a comparison of the predictions made by our model with the results after applying \texttt{sigmaa} to the corresponding incomplete partial structure templates on our test examples in Table \ref{phase_errors}.
Unweighted phase error considers the phase error contribution of all structure factors equally for each example, while weighted phase error weighs the individual phase errors according to the associated figure of merit reported by \texttt{sigmaa}.
To obtain such weighted phase errors for our model predictions, we also applied \texttt{sigmaa} with the same corresponding input parameters as described previously.
The full test set consisted of $64,070$ initial examples, once again with each associated by up to three partial structures with $3-7$ omitted residues, for a total size of $176,556$ examples.
The reported subset consists of $19,436$ (about $11\%$) test set examples with the worst-performing structural alignments of $AF2$-derived partial structure to ground truth according to the Root Mean Square Deviation across all $\alpha$-carbon atoms in the atomic structure.
These examples had such $C\alpha$ RMSD's ranging from $0.564$ to $12.016$ {\AA}, while the full test set had a median RMSD of $0.223$ {\AA}.
Overall, the model predictions (see columns 1,3) show both noticeable improvement and (almost always) decreased variability on all metrics compared to the post-\texttt{sigmaa} baseline (columns 2,4), on both the full test set (leftmost two columns) and the subset of examples for which the initial alignment was relatively poor (rightmost two columns).


\begin{figure*}[!htp]
\includegraphics[width=0.5\textwidth]{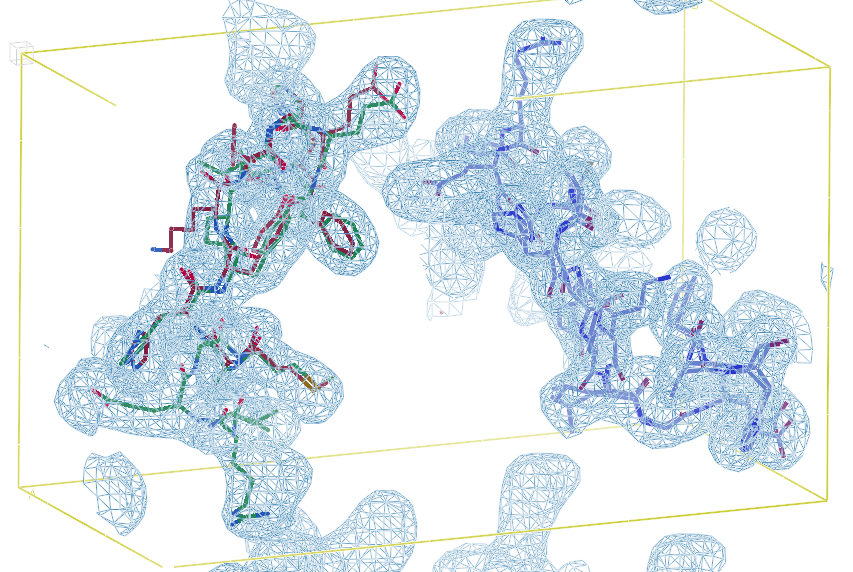} \hfill
\includegraphics[width=0.5\textwidth]{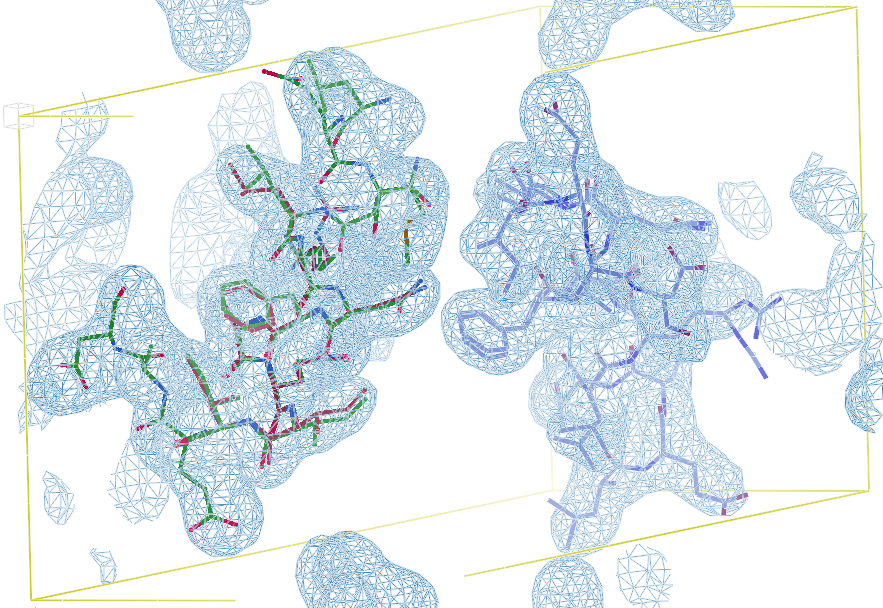} \hfill
\includegraphics[width=0.5\textwidth]{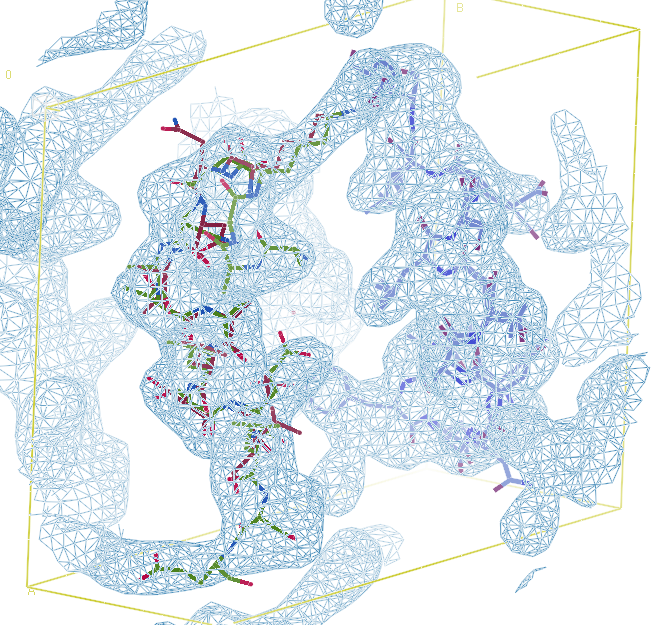} \hfill
\includegraphics[width=0.5\textwidth]{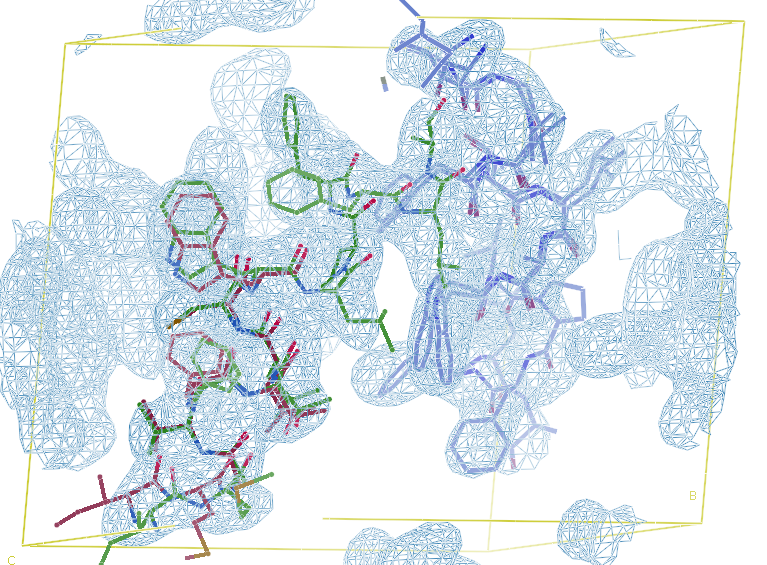} \hfill

\caption{\small Example visualizations of electron density map predictions, shown in blue. The ground truth atomic model is shown in green stick representation, while the partial structure atomic model used to generate the corresponding input template map is shown in red stick representation.}
\label{fig:main} \vspace{-0.3cm}
\end{figure*}

We visualize some predictions in Fig. \ref{fig:main}, with the first row consisting of more typical test set examples and the second consisting of those that belong to the subset of test set examples with the worst partial structure alignment $C\alpha$ RMSD. 
These figures further indicate that not only is the model effective at completing the portions missing from the partial structure template map, but it is also often able to improve regions of poor agreement between the partial structure and the true underlying ground truth.

\begin{figure*}[!htp]
    \vspace{-0.0cm}
    \centering
    \includegraphics[width=0.85\textwidth]{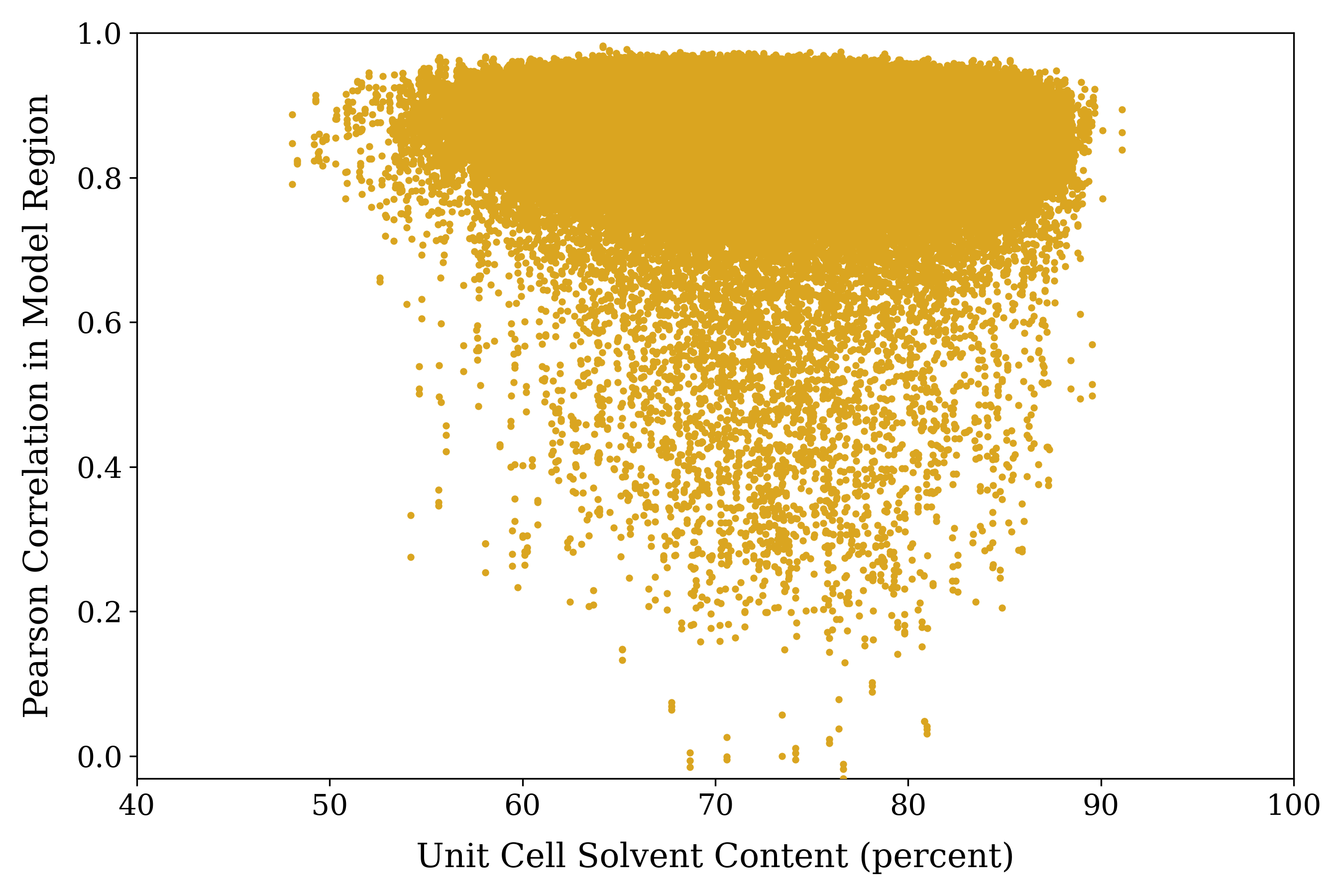} \vspace{-0.1cm}
    \caption{\small Pearson correlations of model predictions with ground truth versus underlying solvent content}
    \label{solvent}
    \vspace{-0.3cm}
\end{figure*}

We also aim to determine what effect, if any, the solvent content of an example has on its difficulty of completion, and so provide a plot of the in-model region Pearson correlations calculated by \texttt{get\_cc\_mtz\_pdb} for our test set model prediction structure factors against the solvent content of the corresponding ground truth in Fig. \ref{solvent}. 
It is clear from the distribution of Pearson correlations that the solvent content of the ground truth unit cell was not an underlying factor in how well the model was able to complete (and fix inaccurate regions of) a partial structure template.

\section{Discussion}\label{sec:discussion}

This work represents a key step in the integration of existing experimental and deep learning-based approaches for protein structural determination.
We have established that our model, \texttt{CrysFormer}, can effectively "complete" small protein fragments with omitted resides, extracted from real predictions taken from the AlphaFold database, when placed in unit cells corresponding to the desired ground truth and used as partial structure template maps alongside Patterson maps obtained from crystallographic data.
Furthermore, the model is often able to greatly improve regions where the template maps were inaccurate.

\subsection{Limitations and Future Work.}
The examples in the dataset we introduce in this work, although closer to true proteins than before, still fall short in realism in several aspects.
Although each example has 30 residues divided across 2 molecules, this is still an unusually small number of residues per unit cell, and our examples still have an unusually high solvent percentage (and thus amount of empty space) in the unit cell.
Thus we are developing a new dataset of examples with fragments consisting of entire protein domains of 50-150 residues in each asymmetric unit.
This introduces another form of variability in our examples, as there will no longer be a constant number of residues in the unit cell across all examples.
We will also use a wider potential range of fractions of residues removed in the training set to further increase model robustness.

Additionally, we want our model to handle more than just one type of internal symmetry at once.
Thus our new dataset will contain examples belonging to 
one of five possible space groups, with up to 4 molecules per unit cell.
We will provide multiple choices of space group and thus unit cell for each original domain example, as yet another form of data augmentation.
This will be necessary to maintain training set size as we will be starting from a much smaller initial set of possible domain examples compared to 15-residue fragments.
We have also not yet considered the effects of bulk solvent or experimental noise when generating our synthetic data examples.

Furthermore, the predictions made by our model after the training run can be incorporated into further ML methods and pipelines.
A prediction can be used as an additional template map input for a subsequent "recycling" iteration of training our model architecture \citep{pan2025reccrysformer}, or its derived structure factors can be used as a set of initial phase estimates for reciprocal-space phasing methods such as phase seeding \citep{Carrozzini:lu5043}.

\section*{Acknowledgements}
This research was funded in part by: The Robert A. Welch Foundation (grant No. C-2118 to G.N.P and A.K.); Rice University (Faculty Initiative award to G.N.P and A.K.); National Science Foundation (NSF), Directorate for Biological Sciences (grant No. 1231306 to G.N.P.);  NSF CAREER (award no. 2145629 to A.K.); a Rice InterDisciplinary Excellence Award (IDEA); an Amazon Research Award; a Microsoft Research Award. The content is solely the responsibility of the authors and does not necessarily represent the official views of the Funders.

\section*{Data availability}
A repository, \url{https://github.com/sciadopitys/CrysFormer\_model\_completion}, contains our CrysFormer model architecture and training and batch generation scripts. It also includes a subset of our dataset generation scripts to generate our Patterson, ground truth, and partial structure maps. Intermediate atomic coordinate files for fragments extracted from the PDB, sufficient to generate ground truth files for the test and training sets, can be downloaded from \url{https://doi.org/10.5281/zenodo.15498745}. Files containing AFDB-derived fragments for partial structure generation can be downloaded from \url{https://doi.org/10.5281/zenodo.15498821}. 

\bibliography{refs} 

\end{document}